\documentclass[aps,pra,twocolumn,showpacs]{revtex4}
\usepackage{graphicx}
\usepackage{graphics}
\begin{document}

\title{Squeezing and entanglement delay using slow light}
\author{Amy Peng, Mattias Johnsson, W. P. Bowen, P. K. Lam, H. -A. Bachor and J. J. Hope}
\affiliation{ARC Center for Quantum-Atom Optics, Faculty of
Science, The Australian National University, ACT 0200, Australia}
\date{\today}

\begin{abstract}
We examine the interaction of a weak probe with $N$ atoms in a
lambda-level configuration under the conditions of
electromagnetically induced transparency (EIT). In contrast to
previous works on EIT, we calculate the output state of the
resultant slowly propagating light field while taking into account
the effects of ground state dephasing and atomic noise for a more
realistic model. In particular, we propose two experiments using
slow light with a nonclassical probe field and show that two
properties of the probe, entanglement and squeezing,
characterizing the quantum state of the probe field, can be
well-preserved throughout the passage.
\end{abstract}

\pacs{42.50.Gy, 03.67.-a}

\maketitle

The coherent and reversible storage of the quantum state of a
light field is an important issue for the realization of many
protocols in quantum information processing. Recently much work
has been done to address this issue by utilizing the phenomenon of
electromagnetically induced transparency
(EIT)~\cite{Harris1997,Fleischhauer2002,Matsko2001,Hau1999,Fleischhauer2000,
Phillips2001,Kash1999,Akamatsu2004,DantanPRA2004,Dantan2004}. In
this paper, we demonstrate that under the conditions of EIT, the
quantum state of the stored light field can be well preserved even
in the presence of dephasing and noise.

In a conventional EIT setup, a strong, coherent field (``control
field") is used to make an otherwise opaque medium transparent
near an atomic resonance. A second, weak field (``probe") with a
restricted bandwidth about this resonance can then propagate
without absorption and with a substantially reduced group velocity
compared to a pulse in vacuum, thus delaying (or effectively
``storing") the light field within the atomic cloud for a duration
equal to the delay time of the light pulse caused by the EIT
medium.

In this paper we concentrate on two representative quantities
characterizing the amount of quantum information of a light field:
squeezing, representing a sub-quantum noise level of fluctuation
in the observable of one beam; and entanglement, where the
sub-quantum noise fluctuation occurs in the correlation between
two beams. We calculate the effect of the atom-light interaction
on each quantity and show that the slowing of the light need not
significantly degrade the information carried. Previous works on
photon storage  have indicated that in the absence of dephasing
between the two ground states of the lambda system, and ignoring
the Langevin noise operators arising from atomic coupling to a
vacuum reservoir, the quantum state of light field is well
preserved after traversing the EIT medium
\cite{Fleischhauer2002,Matsko2001,Lukin2000}. Here we further
highlight the robustness of storage using EIT and show that even
{\it{with}} dephasing and noise taken into account, entanglement
and squeezing of the pulse at the exit of the medium need not
differ significantly from that of the input pulse under
experimentally realizable parameter regimes.

\begin{figure}[h]
\begin{center}
\includegraphics[width=4.5cm,height=2cm]{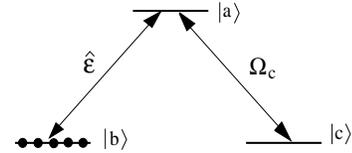}
\caption{$\Lambda$ level structure of the atoms}
 \label{model}
\end{center}
\end{figure}

We follow the model outlined in \cite{Fleischhauer2002} and use a
quasi one-dimensional model, consisting of two co-propagating
beams passing through an optically thick medium of length $L$
consisting of three-level atoms. The atoms have two metastable
lower states $| b \rangle$ and $| c \rangle$ interacting with the
two optical fields $\hat{ \mathcal{E} }(z,t)$ and $\Omega_c$ as
shown in Figure \ref{model}. $\hat{ \mathcal{E} }(z,t)$ is a weak
quantum field that couples the ground state $| b \rangle$ and
excited state $| a\rangle$, and is related to the positive
frequency part of the electric field by
\begin{displaymath}
\hat{E}^{+}(z,t) = \sqrt{ \frac{\hbar \omega_{ab}}{2 \epsilon_0 V}
} \hat{ \mathcal{E} }(z,t) e^{ i \frac{\omega_{ab}}{c}(z-ct) }
\end{displaymath}
where $\omega_{\mu \nu}= (E_{\mu} - E_{\nu})/\hbar$ is the
frequency of the $| \mu \rangle \leftrightarrow | \nu \rangle$
transition. $V$ is the quantization volume of the electromagnetic
field, which is taken to be the interaction volume. The $| c
\rangle \rightarrow | a \rangle$ transition is driven resonantly
by a classical coherent control field with Rabi frequency
$\Omega_c$. We consider the case of copropagating fields to
minimize the effects of Doppler shift.

To perform a quantum analysis of the light-matter interaction it
is useful to introduce locally-averaged atomic operators. Assuming
a length interval $\Delta z$ contains $N_z \gg 1$ atoms over which
the slowly-varying amplitude $\hat{\mathcal{E}}(z,t)$ does not
change much, we can introduce the locally-averaged, slowly-varying
atomic operators
\begin{equation}
\hat{\sigma}_{\mu \nu} (z, t) = \frac{1}{N_z} \sum_{z_j \in N_z}
\hat{\sigma}_{\mu \nu}^{j} (t) e^{i \frac{\omega_{\mu \nu}}{c}(z -
ct)} \label{localatom}
\end{equation}
where $\hat{\sigma}^j_{\mu \nu}(t) = |\mu^j (t) \rangle \langle
\nu^j (t) |$ for the $j$th atom.

Going to the continuum limit, the interaction Hamiltonian can be
written in terms of the locally-averaged atomic operators as
\begin{equation}
\hat{\mathcal{H} } = - \int \frac{N \hbar}{L} [ g
\hat{\sigma}_{ab}(z,t) \hat{\mathcal{E}}(z,t) + \Omega_c(z,t)
\hat{\sigma}_{ac}(z,t) + H.c. ] dz \label{hamiltonian}
\end{equation}
where $g = d_{ba} \sqrt{ \omega_{ab} /2 \epsilon_0 V \hbar}$ is
the atom-field coupling constant, $d_{ba}$ is the atomic dipole
moment for the $|b\rangle \leftrightarrow |a\rangle$ transition
and $L$ is the cell length. The equations of motion are then given
by
\begin{eqnarray}
\dot{\hat{\sigma}}_{bb} & = & \gamma_b \hat{\sigma}_{aa} +
\gamma_{bc} (\hat{\sigma}_{cc} - \hat{\sigma}_{bb}) - i g
\hat{\mathcal{E}} \hat{\sigma}_{ab} + i g^{\ast}
\hat{\mathcal{E}}^{\dagger} \hat{\sigma}_{ba} + \hat{F}_{bb} \nonumber \\
\dot{\hat{\sigma}}_{cc} & = & \gamma_c \hat{\sigma}_{aa} +
\gamma_{bc}(\hat{\sigma}_{bb} - \hat{\sigma}_{cc}) - i \Omega_c
\hat{\sigma}_{ac} + i \Omega_c^{\ast} \hat{\sigma}_{ca} +
\hat{F}_{cc} \nonumber \\
\dot{\hat{\sigma}}_{ba} & = & -\gamma_{ba} \hat{\sigma}_{ba} + i g
\hat{\mathcal{E}} (\hat{\sigma}_{bb} - \hat{\sigma}_{aa} ) + i
\Omega_c \hat{\sigma}_{bc} + \hat{F}_{ba} \nonumber \\
\dot{\hat{\sigma}}_{bc} & = & -\gamma_{bc} \hat{\sigma}_{bc} - i g
\hat{\mathcal{E}} \hat{\sigma}_{ac} + i \Omega_c^{\ast}
\hat{\sigma}_{ba} + \hat{F}_{bc} \nonumber \\
\dot{\hat{\sigma}}_{ac} & = & -\gamma_{ac} \hat{\sigma}_{ac} - i
g^{\ast} \hat{\mathcal{E}}^{\dagger} \hat{\sigma}_{bc} + i
\Omega_c^{\ast} (\hat{\sigma}_{aa} - \hat{\sigma}_{cc}) +
\hat{F}_{ac} \nonumber \\
\left( \frac{\partial}{\partial t} \right. & + & \left. c
\frac{\partial}{\partial z} \right) \hat{\mathcal{E}}  =  i
g^{\ast} N \hat{\sigma}_{ba} \label{motion2}
\end{eqnarray}
where we have included the decays of the atomic dipole operators
$\gamma_{\mu}$ and the associated Langevin noise operators which
describe the effect of spontaneous decay caused by the coupling of
atoms to all the vacuum field modes. The random decay process adds
noise to individual atomic operators represented by the single
atom Langevin noise operators $\hat{F}_{\mu \nu}^{i} (t)$. The
continuous Langevin noise operators are related to the single-atom
noise operators by the same relation as equation (\ref{localatom})
\begin{displaymath}
\hat{F}_{\mu \nu} (z,t) = \frac{1}{N_z} \sum_{z_j \in N_z}
\hat{F}_{\mu \nu}^i (t) e^{i \frac{\omega_{\mu \nu}}{c} (z - c t)
}.
\end{displaymath}
The decay rate $\gamma_{bc}$ of the coherence between the two
ground states is of critical importance, and arises chiefly from
atomic collisions and atoms drifting out of the interaction
region. For rubidium vapor cells with a buffer gas, typically
$\gamma_{bc} \approx 1$ kHz although it is possible to attain
$\gamma_{bc} \approx 160$ Hz \cite{Kash1999}.

In order to solve the propagation equations we make the usual
assumption that the quantum field intensity is much less than that
of the classical control field
$\Omega_c$~\cite{Fleischhauer2002,Scully}. Assuming all the atoms
are initially in the state $| b \rangle$ we can solve equations
(\ref{motion2}) perturbatively to first order in
$g\hat{{\mathcal{E}}}/\Omega_c$ to obtain a set of three closed
equations
\begin{eqnarray}
\dot{\hat{\sigma}}_{ba} & = & - \gamma_{ba} \hat{\sigma}_{ba} + i
g \hat{\mathcal{E}} + i \Omega_c \hat{\sigma}_{bc} + \hat{F}_{ba}
\label{sigmaba} \\
\dot{\hat{\sigma}}_{bc} & = & - \gamma_{bc} \hat{\sigma}_{bc} + i
\Omega_c^{\ast} \hat{\sigma}_{ba} + \hat{F}_{bc} \label{sigmabc}
\\
\left( \frac{\partial}{\partial t} \right. & + & \left. c
\frac{\partial}{\partial z} \right) \hat{\mathcal{E}}  =  i
g^{\ast} N \hat{\sigma}_{ba}. \label{field}
\end{eqnarray}
To solve these equations we Fourier transform to the frequency
domain via
\begin{displaymath}
\tilde{F}(z,\omega) = \frac{1}{\sqrt{2 \pi}}
\int_{-\infty}^{\infty} \hat{F}(z,t) e^{i \omega t} dt,
\end{displaymath}
where $\omega=0$ corresponds to the carrier frequency
$\omega_{ab}$ in the interaction picture. We solve (\ref{sigmaba})
and (\ref{sigmabc}) for $\hat{\sigma}_{ba}$ in terms of
$\hat{\mathcal{E}}$, substitute into (\ref{field}), and perform
formal integration over $z$ to obtain the field at the exit of the
cell after interaction with the EIT medium. We find
\begin{eqnarray}
 \tilde{\mathcal{E}} (L, \omega) &=&  e^{- \Lambda(\omega) L}
\tilde{\mathcal{E}}(0,\omega)
  + \frac{g^{\ast} N}{c}  \int_0^L  e^{-\Lambda(\omega)(L-s)} \nonumber \\
  &\hspace{-0.4cm}\times & \hspace{-0.4cm} \left[
\frac{-\Omega_c \tilde{F}_{bc}(s,\omega) + (\omega + i
\gamma_{bc}) \tilde{F}_{ba}(s,\omega)}{ (\gamma_{ab} - i
\omega)(\gamma_{bc} - i \omega) + | \Omega_c |^2 } \right] ds
 \label{solution}
\end{eqnarray}
with
\begin{equation}
\Lambda(\omega) = \frac{ \frac{|g|^2 N}{c} (\gamma_{bc} - i
\omega)}{(\gamma_{ba} - i \omega)(\gamma_{bc} - i \omega) +
|\Omega_c|^2} - \frac{i \omega}{c} \label{suscep}
\end{equation}
Equation (\ref{solution}) can be interpreted as follows: The
amplitude of the field operator is attenuated and phase shifted
according to the function $\Lambda(\omega)$, the values of which
depend on the actual frequency component of the field. Expanding
$\Lambda(\omega) L$ about the carrier frequency $\omega=0$ gives
\begin{equation}
\Lambda(\omega) L = K L - \frac{i \omega L}{v_g} +
\frac{\omega^2}{\delta \omega^2} + O(|\omega|^3)
\label{approxsuscep}
\end{equation}
where
\begin{eqnarray}
K & = & \frac{N |g|^2 \gamma_{bc}}{c(\gamma_{ba} \gamma_{bc} +
|\Omega_c|^2)} \nonumber \\
v_g & = & \frac{c}{1 + \frac{N |g|^2 (|\Omega_c|^2 -
\gamma_{bc}^2)}{(\gamma_{ba} \gamma_{bc} + |\Omega_c|^2)^2}}
\nonumber \\
\delta \omega^2 & = & \frac{ c(\gamma_{ba} \gamma_{bc} +
|\Omega_c|^2)^3}{N |g|^2 L (|\Omega_c|^2(2 \gamma_{bc} +
\gamma_{ba})-\gamma_{bc}^3)} \nonumber
\end{eqnarray}
The zeroth order term represents attenuation due to the finite
coherence lifetime between the two ground states which is
proportional to the dephasing rate $\gamma_{bc}$. Note that it
also prevents perfect transparency even at resonance and will
probably be the ultimate limitation on storage using EIT type
techniques. However, it is possibly to significantly reduce
$\gamma_{bc}$ by using Bose-Einstein condensates, for example.

The first order term in $\omega$ describes a modification of the group
velocity from
\begin{displaymath}
c \rightarrow v_g = \frac{c}{ 1 + \frac{N |g|^2 (|\Omega_c|^2 -
\gamma_{bc}^2)}{(\gamma_{ba} \gamma_{bc} + |\Omega_c|^2)^2}}
\end{displaymath}
which can be many orders of magnitudes smaller than $c$. For
$|\Omega_c| \gtrsim 2 \gamma_{bc}$ the effect of the nonzero
$\gamma_{bc}$ on the group velocity is hardly noticeable. The
nonzero dephasing rate does, however, prevent $v_g$ from reaching
zero. In fact, the group velocity increases towards $c$ as
$|\Omega_c|$ decreases below $\sim 2 \gamma_{bc}$, although
significant absorption due to the breakdown of EIT will have
already occurred once $|\Omega_c|$ becomes comparable to
$\gamma_{bc}$ and thus it is difficult to speak meaningfully of a
group velocity.

The quadratic term in (\ref{approxsuscep}) represents absorption
of the high frequency components that fall outside the EIT
transparency window $\delta \omega$ \cite{Fleischhauer2002}. This
justifies the Taylor's expansion of $\Lambda(\omega)$, since to
prevent significant pulse distortion, one would choose the
bandwidth of the input pulse $\Delta \omega$ to satisfy $\Delta
\omega / \delta \omega \ll 1$. For $|\Omega_c| \leq \gamma_{ab}$
and small $\gamma_{bc}$, $\delta \omega$ is approximately a
linearly decreasing function of $\gamma_{bc}$, a tool which could
be used experimentally to estimate the value of $\gamma_{bc}$.

Returning to (\ref{solution}), the second term on the right
hand side represents vacuum noise added to the probe field
as it interacts with the atoms. We now determine under
what conditions this noise contribution is small.

As we are interested in whether the quantum properties of the
input field survive the passage throught the EIT setup, we
consider two quantities of interest: one relating to entanglement
and the other to squeezing. Here we propose some possible
experiments that could be performed to discern the effect of slow
light on entanglement and squeezing, and show how the presence of
the dephasing and decay mechanisms that we have included in our
model manifest themselves in the output beam for each type of
experiment.

\begin{figure}
\begin{center}
\includegraphics[width=6cm, height = 2.5cm]{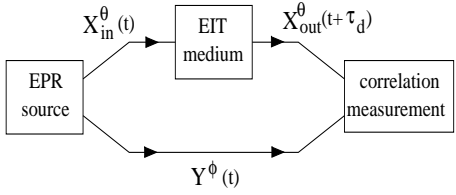}
\caption{Setup to measure whether entanglement of slowed light is
preserved. $\hat{X}$ and $\hat{Y}$ are two entangled beams where
$\hat{X}$ passes through an EIT medium and is consequently delayed
by time $\tau_d$ while $\hat{Y}$ travels in vacuum. The angle
$\theta$ and $\phi$ denotes the specific quadrature to be
interrogated.} \label{entangle}
\end{center}
\end{figure}
To investigate the effect of slow light on squeezing, we consider
an experiment where the amount of squeezing before and after
passing through the EIT medium are compared \cite{Akamatsu2004}.
For an entanglement measurement, we propose a setup shown in
Figure \ref{entangle}. Starting with a pair of entangled beams,
which could be produced, for example, from parametric down
conversion, one beam passes through an EIT medium of length L and
is consequently delayed by a time $\tau_d$ while the other beam
passes through the same length of vacuum. At the output, we
perform a measurement to quantify the degree of entanglement
between the field that was slowed and the field that was not.

To quantify squeezing and entanglement, it is usual to consider
the field quadrature operator which has a Fourier transform in the
temporal domain of
\begin{equation}
\tilde{X}^{\theta}_{out}(\omega) = \hat{\mathcal{E}}(L, \omega)
e^{i \theta} + \tilde{\mathcal{E}}^{\dagger}(L, -\omega) e^{-i
\theta} \label{quadrature}.
\end{equation}
Using the solution (\ref{solution}), the output quadrature
operator is related to the input via the relation
\begin{eqnarray}
\tilde{X}_{out}^{\theta}(\omega) & = &
\tilde{X}_{in}^{\theta}(\omega)
e^{- \Lambda(\omega) L} \label{quadratureout} \\
-\frac{N g}{c} & \int_0^L & e^{-\Lambda(\omega)(L-s)} \nonumber \\
& \times & \frac{\Omega_c \tilde{F}_{bc}(s,\omega)e^{i \theta} +
\Omega_c^{\ast} \tilde{F}_{bc}^{\dagger}(s, -\omega)e^{-i
\theta}}{(\gamma_{ba} -
i \omega)(\gamma_{bc} - i \omega) + |\Omega_c|^2} ds \nonumber \\
+ \frac{N g}{c} & \int_0^L & e^{-\Lambda(\omega) (L-s)} \nonumber
\\
& \times & (\omega + i \gamma_{bc})
\frac{\tilde{F}_{ba}(s,\omega)e^{i \theta} -
\tilde{F}_{ba}^{\dagger}(s, -\omega)e^{-i \theta}}{(\gamma_{ba} -
i \omega)(\gamma_{bc} - i \omega) + |\Omega_c|^2} ds \nonumber
\end{eqnarray}
where we take $g$ to be real for simplicity.

To calculate the degradation of squeezing after passing through
the EIT medium, we calculate the output squeezing flux spectrum
defined by
\begin{equation}
\mathcal{S}_{out} (\omega) \; \delta(\omega + \omega') =
\frac{c}{L} \langle \tilde{X}_{out}(\omega)
\tilde{X}_{out}(\omega') \rangle \label{spectrum}
\end{equation}

In order to compute (\ref{spectrum}) using (\ref{quadratureout})
it is necessary to calculate the correlation functions of the
Langevin noises involved. These can be derived using the
generalized Einstein relations. In the space-time domain, the
generalized Einstein relation for the single atom operators can be
written as \cite{Tannoudji}
\begin{eqnarray}
\langle \hat{F}^i_{\mu \nu}(t_1) \hat{F}^j_{\alpha \beta}(t_2)
\rangle & = & \langle \mathcal{D}(\hat{\sigma}_{\mu \nu}^i
\hat{\sigma}_{\alpha \beta}^i ) - \mathcal{D}(\hat{\sigma}_{\mu
\nu}^i) \hat{\sigma}_{\alpha \beta}^i - \hat{\sigma}_{\mu \nu}^i
\mathcal{D}(\hat{\sigma}_{\alpha \beta}^i) \rangle \nonumber \\
& \times & \delta(t_1 - t_2) \delta_{ij} \label{Einstein}
\end{eqnarray}
where the notation $\mathcal{D}(\hat{\sigma}_{\mu \nu}^i)$ denotes
the deterministic part of the Heisenberg equation of motion for
$\hat{\sigma}_{\mu \nu}^i$, that is the equation for
$\dot{\hat{\sigma}}_{\mu \nu}^i$ with the Langevin noise terms
omitted. The Dirac delta function in (\ref{Einstein}) represents
the short memory of the vacuum reservoir modes while the Kronecker
delta occurs because we assume each atom couples only to its own
reservoir.

Using the definition of the locally-averaged Langevin force
operators in terms of their single atom counterpart, we derive the
following for the nonzero correlations of the continuous Langevin
correlations. After transforming to the frequency domain, these
are
\begin{eqnarray}
\langle \tilde{F}_{ba}(z_1, \omega_1)
\tilde{F}_{ab}^{\dagger}(z_2, \omega_2) \rangle & = &
\frac{\delta(z_1 - z_2) \delta(\omega_1 +
\omega_2)}{n \mathcal{A}} \label{baab} \\
\times (\gamma_{ba} \langle \hat{\sigma}_{aa} \rangle & + & 2
\gamma_{ba} \langle \hat{\sigma}_{bb} \rangle - \gamma_{bc}
\langle \hat{\sigma}_{bb} - \hat{\sigma}_{cc} \rangle )
\nonumber \\
\langle \tilde{F}_{ba}^{\dagger}(z_1,\omega_1)
\tilde{F}_{bc}(z_2,\omega_2) \rangle & = & \frac{\delta(z_1 -
z_2) \delta(\omega_1 + \omega_2)}{n \mathcal{A} } \label{abbc} \\
& \times & \gamma_{bc} \langle \hat{\sigma}_{ac} \rangle
\nonumber \\
\langle \tilde{F}_{bc}^{\dagger}(z_1,\omega_1)
\tilde{F}_{ba}(z_2,\omega_2) \rangle & = & \frac{\delta(z_1-z_2)
\delta(\omega_1 + \omega_2)}{n \mathcal{A} } \label{cbba} \\
& \times & \gamma_{bc} \langle \hat{\sigma}_{ca} \rangle \nonumber \\
\langle \tilde{F}_{bc}(z_1,\omega_1)
\tilde{F}_{bc}^{\dagger}(z_2,\omega_2) \rangle & = &
\frac{\delta(z_1-z_2) \delta(\omega_1 + \omega_2)}{n \mathcal{A}}
\label{bccb} \\
& \times & (\gamma_{ba} \langle \hat{\sigma}_{aa} \rangle +
\gamma_{bc} \langle \hat{\sigma}_{cc} + \hat{\sigma}_{bb} \rangle
) \nonumber \\
\langle \tilde{F}^{\dagger}_{bc}(z_1,\omega_1)
\tilde{F}_{bc}(z_2,\omega_2) \rangle & = & \frac{\delta(z_1-z_2)
\delta(\omega_1 + \omega_2)}{n \mathcal{A}} \label{cbbc} \\
& \times & (\gamma_{ba} \langle \hat{\sigma}_{aa} \rangle +
\gamma_{bc} \langle \hat{\sigma}_{cc} + \hat{\sigma}_{bb} \rangle
) \nonumber
\end{eqnarray}
where $n$ is the atomic density, $\mathcal{A}$ is the cross
section area of the beam and we have taken $\gamma_b = \gamma_c =
\gamma_{ba} = \gamma_{ca}$ in equation (\ref{motion2}) for
convenience.

We substitute (\ref{quadratureout}) into (\ref{spectrum}) and
simplify using (\ref{baab}) - (\ref{cbbc}). In accordance with the
weak probe assumption we also set $\langle \hat{\sigma}_{aa}
\rangle \approx \langle \hat{\sigma}_{cc} \rangle \approx \langle
\hat{\sigma}_{ac} \rangle \approx 0$. The output squeezing
spectrum (normalized with shot noise at 1) is
\begin{eqnarray}
\mathcal{S}_{out}(\omega)  & = & \mathcal{S}_{in}(\omega) e^{-2
\Re\{ \Lambda(\omega) \} L}  \label{outspectrum} \\
& + & \frac{N |g|^2}{c} \left[ \frac{1 - e^{-2 \Re \{
\Lambda(\omega) \} L}}{2 \Re \{ \Lambda(\omega) \}} \right]
\nonumber \\
& \times & \frac{ (\omega^2 + \gamma_{bc}^2)(2 \gamma_{ba} -
\gamma_{bc}) + 2 |\Omega_c|^2 \gamma_{bc}}{|(\gamma_{ba} - i
\omega)(\gamma_{bc} - i \omega) + |\Omega_c|^2 |^2}  \nonumber.
\end{eqnarray}

Before discussing how much the squeezing in the probe beam
degrades due to the slow light propagation, we first consider how
the entanglement between two beams degrades after one of the
fields passes through an EIT medium. If we define the difference
operator between the quadratures of the two beams
$\hat{X}_{in}^{\theta}(t)$ and $\hat{Y}_{in}^{\phi}(t)$ as
\begin{equation}
\hat{Z}_{in}(\theta, \phi, t) = \hat{X}_{in}^{\theta}(t)
-\hat{Y}_{in}^{\phi}(t) \label{differencequad}
\end{equation}
then the two beams are said to be entangled when {\it both} of the
combinations involving non-commuting observables for one of the
beams $\hat{Z}_{in}(\theta, \phi, t)$ and
\mbox{$\hat{Z}_{in}(\theta + \pi/2, \phi - \pi/2,t)$} are squeezed
\cite{Duan2000, Bowen2003}. Note that squeezing in both is
required to constitute an Einstein-Podolsky-Rosen (EPR) paradox
\cite{Einstein1935}, and that squeezing in just one variable is
insufficient.

At the output of the EIT medium, we account for the effects of
slow light by looking for squeezing in the time adjusted variable
\begin{equation}
\hat{Z}_{out}(\theta, \phi,t) = \hat{X}_{out}^{\theta}(t + \tau_d)
- \hat{Y}_{out}^{\phi}(t) \label{entanglevar}
\end{equation}
where $\tau_d = L(1/v_g - 1/c)$ is the delay due to the EIT effect
compared to light that had travelled distance $L$ in vacuum.
Again, for a nonclassical correlation between the two beams, we
require squeezing in $\hat{Z}_{out}(\theta, \phi, t)$ and
$\hat{Z}_{out}(\theta + \pi/2, \phi - \pi/2,t)$. Fourier
transforming equation (\ref{entanglevar}) we obtain \mbox{
$\tilde{Z}_{out}(\theta, \phi, \omega) =
\tilde{X}_{out}^{\theta}(\omega) e^{-i \omega \tau_d} -
\tilde{Y}^{\phi}_{out}(\omega)$} with $\tilde{Y}_{out}^{\phi} =
\tilde{Y}_{in}^{\phi} e^{i \omega L/c }$, the phase shift arising
from free evolution of the light field that has propagated through
a distance $L$ in vacuum at speed $c$.

Defining the flux of entanglement spectrum for
$\tilde{Z}_{out}(\theta, \phi, \omega)$ in a way analogous to
equation (\ref{spectrum})
\begin{equation}
\mathcal{A}_{out}(\theta, \phi, \omega) \; \delta(\omega +
\omega') = \frac{c}{L} \langle \tilde{Z}_{out}(\theta, \phi,
\omega) \tilde{Z}_{out}(\theta, \phi, \omega') \rangle
\label{entanglespectrum}
\end{equation}
the entanglement criteria described above (called Duan's
inseparability criteria \cite{Duan2000}) can be recast using
Duan's inseparability measure $\mathcal{I}(\omega)$ into the form
\cite{Duan2000, Bowen2003}
\begin{equation}
\mathcal{I}(\omega) = \sqrt{\mathcal{A}_{out}(\theta, \phi,
\omega) \mathcal{A}_{out}(\theta + \pi/2, \phi - \pi/2,\omega)} <
1 \label{EPRcriteria}
\end{equation}

Evaluating the right hand side of (\ref{entanglespectrum}) using
equations (\ref{quadratureout}) and (\ref{baab}) - (\ref{cbbc}) we
get
\begin{eqnarray}
& \frac{L}{c} & \delta(\omega + \omega') \mathcal{A}_{out}(\theta,
\phi, \omega) = \langle \tilde{X}^{\theta}_{in}(\omega)
\tilde{X}^{\theta}_{in}(\omega') \rangle e^{-2 \Re \{
\Lambda(\omega) \} L} \nonumber \\
& - & \langle \tilde{X}_{in}(\omega)^{\theta}
\tilde{Y}_{in}^{\phi}(\omega') \rangle e^{-[\Lambda(\omega) +
\frac{i \omega}{v_g} ]L} \nonumber \\
& - & \langle \tilde{Y}^{\phi}_{in}(\omega)
\tilde{X}^{\theta}_{in}(\omega') \rangle
e^{-[\Lambda(-\omega)-\frac{i \omega}{v_g}]L} \nonumber \\
& + & \langle \tilde{Y}^{\phi}_{in}(\omega)
\tilde{Y}^{\phi}_{in}(\omega')
\rangle \label{entangleout} \\
& + & \delta(\omega + \omega') \frac{N |g|^2}{c} \left[ \frac{ 1 -
e^{-2 \Re \{ \Lambda(\omega) \} L}}{2 \Re \{ \Lambda(\omega) \} }
\right]  \nonumber \\
& \times & \frac{(\omega^2 + \gamma_{bc}^2)(2 \gamma_{ba} -
\gamma_{bc}) + 2 |\Omega_c|^2 \gamma_{bc}}{|(\gamma_{ba} - i
\omega)(\gamma_{bc} - i \omega) + |\Omega_c|^2|^2} \nonumber
\end{eqnarray}
Note that the noise addition represented by the last term in
(\ref{entangleout}) is identical to the one that appeared in the
output squeezing spectrum, a clear consequence of our choice of
entanglement measure. Also from this term, we see that the
destructive effects are independent of the particular quadrature
considered. In other words, the imperfections in the setup
adversely affect the phase and amplitude correlations by the same
amount.

For a clearer insight into how the entanglement is affected, it is
instructive to utilize the Taylor's expansion for
$\Lambda(\omega)$ in equation (\ref{approxsuscep}) and throw away
the quadratic and higher order terms on the grounds that the probe
bandwidth is well inside the transparency window. Then under the
parameter regime when the absorption is low ($KL \ll 1$), the
entanglement at the output is
\begin{eqnarray}
\mathcal{A}_{out}(\theta, \phi, \omega) & \approx &
\mathcal{A}_{in}(\theta, \phi, \omega) e^{-K L} + \frac{N
|g|^2}{c} \left[ \frac{ 1 - e^{-2 K L}}{2 K} \right] \nonumber
 \\
&\times& \frac{(\omega^2 + \gamma_{bc}^2)(2 \gamma_{ba} -
\gamma_{bc}) + 2 |\Omega_c|^2 \gamma_{bc}}{|(\gamma_{ba} - i
\omega) (\gamma_{bc} - i \omega) + |\Omega_c|^2|^2}.
\label{entangleapprox}
\end{eqnarray}

As an indication of the amount of degradation in squeezing and
entanglement, we consider a cell of length 3.5 cm with atomic
density of $ 1 \times 10^{12}$ atoms per cm$^3$, $\gamma_{ba} = 6
\pi$ MHz, $\gamma_{bc} = 10$ Hz, and $\Omega_c = 30 \pi$ MHz. The
modified group velocity is 3100 ms$^{-1}$ corresponding to a delay
time of $11 \, \mu s$ and a transparency window of 6.5 MHz. For
normalized noise variance and choosing Duan's inseparability
measure as defined in equation (\ref{EPRcriteria}) to be initially
0.4 at 1 MHz (1.0 is the standard quantum limit), the output
variance characterizing squeezing retained in the slowed light is
0.43, and the output entanglement is 0.45. In contrast, keeping
all other parameters the same but choosing $\gamma_{bc} = 5$ kHz,
the normalized noise variance and entanglement measure at the
output are 0.49 and 0.53 respectively, with no significant changes
in group velocity or transparency window. Thus we see that
preservation of quantum properties relies crucially on having
small values of $\gamma_{bc}$ as well as the probe bandwidth
staying well inside the transparency window, as can be seen from
the presence of $\omega^2$ type terms in equations
(\ref{outspectrum}) and (\ref{entangleout}). Mechanisms for
reducing $\gamma_{bc}$ via buffer gas in vapor cells have been
investigated experimentally \cite{Kash1999,Brandt1997}.

In conclusion, we have demonstrated that both entanglement and
squeezing of the probe field can be almost perfectly preserved
under slow light setup, even when quantum noise due to atom-light
interactions are taken into account, provided the ground state
decoherence rate $\gamma_{bc}$ is sufficiently small. In many
experimental situations, however, there may be other mechanisms,
for example, coupling to states outside the $\Lambda$ system,
which are the dominant contribution to losses in EIT
\cite{Deng2001}. Also, we have only considered the slow light
scenario while the key to true storage of light using EIT relies
on dynamically controlling $\Omega_c$ as outlined in
\cite{Fleischhauer2002}. Nevertheless, our results still underline
the robustness of quantum information delay using an atomic lambda
system, allowing the possibility of using slow light for squeezing
or entanglement experiments.

We acknowledge helpful discussions with John Close and funding
from the Australian Research Council.

\end{document}